\documentclass[%
 reprint,
superscriptaddress,
 amsmath,amssymb,
 aps,
]{revtex4-2}

\usepackage{graphicx}
\usepackage{dcolumn}
\usepackage{bm}
\usepackage{upgreek}
\usepackage{xcolor}
\usepackage[version=3]{mhchem}
\usepackage{mathtools, nccmath}
\usepackage{lipsum}

\begin{document}
\title{Surface-Mediated Molecular Transport of a Lipophilic Fluorescent Probe in Polydisperse Oil-in-Water Emulsions}

\author{Marius~R.~Bittermann}
\email{m.r.bittermann@uva.nl}
\affiliation{Van der Waals-Zeeman Institute, IoP, University of Amsterdam, Science Park 904, 1098 XH Amsterdam, Netherlands.}
\author{Tatiana~I.~Morozova}
\email{morozova@ill.fr}
\affiliation{Institut Laue-Langevin, 71 Avenue des Martyrs, Grenoble 38042, France}
\author{Santiago~F.~Velandia}
\affiliation{Van der Waals-Zeeman Institute, IoP, University of Amsterdam, Science Park 904, 1098 XH Amsterdam, Netherlands.}
\author{Elham~Mirzahossein}
\affiliation{Van der Waals-Zeeman Institute, IoP, University of Amsterdam, Science Park 904, 1098 XH Amsterdam, Netherlands.}
\author{Antoine~Deblais}
\affiliation{Van der Waals-Zeeman Institute, IoP, University of Amsterdam, Science Park 904, 1098 XH Amsterdam, Netherlands.}
\author{Sander~Woutersen}
\affiliation{Van ’t Hoff Institute for Molecular Sciences, University of Amsterdam, Science Park 904, 1098 XH Amsterdam, Netherlands.}
\author{Daniel~Bonn}
\email{d.bonn@uva.nl}
\affiliation{Van der Waals-Zeeman Institute, IoP, University of Amsterdam, Science Park 904, 1098 XH Amsterdam, Netherlands.}    

\begin{abstract}
  Emulsions often act as carriers for water-insoluble solutes that are delivered to a specific target. The molecular transport of solutes in emulsions can be facilitated by surfactants and is often limited by diffusion through the continuous phase. We here investigate this transport on a molecular scale by using a lipophilic molecular rotor as a proxy for solutes. Using fluorescence lifetime microscopy we track the transport of these molecules from the continuous phase towards the dispersed phase in polydisperse oil-in-water emulsions. We show that this transport comprises two timescales, which  vary significantly with droplet size and surfactant concentration, and, depending on the type of surfactant used, can be limited either by transport across the oil-water interface, or by diffusion through the continuous phase. By studying the time-resolved fluorescence of the fluorophore, accompanied by molecular dynamics simulations, we demonstrate how the rate of transport observed on a macroscopic scale can be explained in terms of the local environment that the probe molecules are exposed to.
\end{abstract}

\maketitle

\section{Introduction}
In its simplest form, an emulsion is a surfactant-stabilized mixture of immiscible liquids, in which one phase is dispersed in the other \cite{leal2007}. One increasingly popular application of emulsions is as delivery systems for bioactive solutes, such as for drugs \cite{buyukozturk2010,mcclements2012,wadhwa2012} or for functional food ingredients \cite{mcclements2007}. Considering drug delivery, one has to keep in mind that most newly discovered drugs are lipophilic, i.e. poorly soluble in water \cite{porter2007lipids, waring2010lipophilicity, arnott2012influence}. For such drugs, emulsions tend to be promising delivery agents, given that their oil phase solubilizes lipophilic drugs while they retain a high bioavailability \cite{wadhwa2012}.
As an example, in a recent study oil-in-water emulsions were shown to be promising delivering agents for the topical delivery of the lipophilic drug bifonazole \cite{hiranphinyophat2021particle}.
From a thermodynamic viewpoint, emulsions are complex; they are in a metastable state stabilized by surfactants, molecules that adsorb to the interface between the oil and aqueous phases.  
When ageing, emulsions tend to destabilize, which involves mechanisms such as flocculation, creaming or coalescence. Emulsion ageing is also accompanied by a material flow that can be composed of the dispersed phase itself, in a process referred to as Ostwald ripening, \cite{kabalnov1987,taylor1998} or of solutes being exchanged between the phases \cite{fletcher1987,courtois2009,chen2012,skhiri2012,gruner2016,etienne2018}. For emulsions that act as delivery systems for solutes, a good understanding of the dynamics of such molecular transports is crucial, given that the solutes partition between the phases of the emulsions and eventually have to be delivered to a target.
Baret and coworkers have shown that for monodisperse water-in-oil emulsions, the exchange of solutes, poorly soluble in the continuous phase, is mediated by micelles and limited by diffusion through the continuous phase \cite{gruner2016}. The diffusive process was demonstrated to be faster with increasing surfactant concentration but slower with increasing spacing between the droplets.\\
Here, we investigate the transport of a lipophilic molecule solubilized in micelles in the continuous phase to the interior of the droplets in a polydisperse oil-in-water emulsion. As a model system for lipophilic solutes, we use the dye molecule BODIPY-C12, a popular probe to study membranes on the nanoscale \cite{kuimova2012mapping,wu2013,lopez2014molecular}.
By tracking BODIPY-C12 in these emulsions we find that the molecular transport from the continuous phase into the oil droplets is a two-step process; the slow depletion of BODIPY-C12 from the continuous phase is followed by dye exchange between the oil droplets. 
The rate of transport shows a dependence on surfactant concentration and droplet size, which we explain using simple models based on permeability theory and diffusion.
We find that for some surfactants the partitioning of the solute molecules at the surface of droplets can become the limiting step, where the solute exchange is slowed down to a timescale of days.
Surprisingly, in our system the dynamics of the transport can neither be explained by the surfactant polarity (i.e., the hydrophilic–lipophilic balance), nor electrostatic interactions.
Instead, analysis of the time-resolved fluorescence of the fluorophore suggests that the retention at the oil-water interface is due to interactions on a molecular level, in particular due to the mobility of BODIPY-C12 in its local environment and the size of the surfactant molecules.
To push further our understanding of the process, we perform coarse-grained molecular dynamics simulations that confirm the experimental observations. Additionally, simulation results suggest that parameters and interactions governing this surface-mediated molecular transport are the interactions between the head groups of the dye and surfactant molecules.

\section{Methods}
\subsection{Emulsion Preparation}
\begin{figure}[htb]
\centering
\includegraphics[width=\columnwidth]{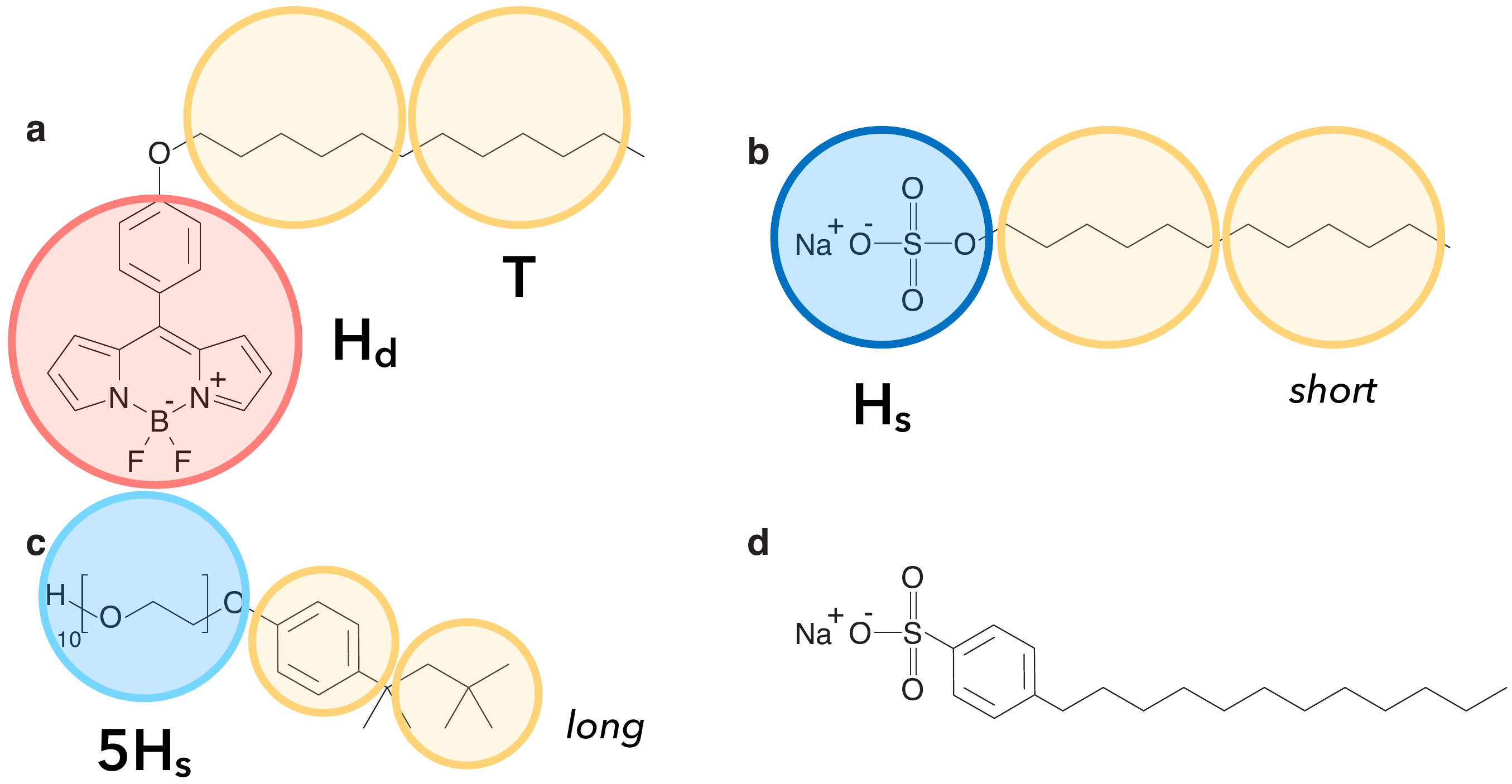}
\caption{Chemical structures of the molecular rotor BODIPY-C12 (a), and the surfactants SDS (b), TX-100 (c) and SDBS (d). Schematic representation of the corresponding simulation models and the bead types are shown as well. Note that all beads have the same size in simulations.} 
\label{fig:structures}
\end{figure}

Emulsions were prepared by dispersing viscous polydimethylsiloxane silicone oil (500 cst, from Sigma-Aldrich) in aqueous sodium dodecyl sulfate (SDS,  $\geq$ 99.0\%, from Sigma), Sodium dodecylbenzenesulfonate (SDBS, technical grade, from Sigma-Aldrich), and \textit{t}-octylphenoxypolyethoxyethanol (TX-100, laboratory grade, from Sigma-Aldrich) solutions.
We chose an oil volume fraction of $80\%$ and surfactant concentrations of $1\, \mathrm{wt}\%$ and $2\, \mathrm{wt}\%$, which are well above the critical micelle concentrations ($\approx4$ and $\approx8$ $\times$ cmc, respectively) \cite{mukerjee1971critical}.
Using a Silverson high-shear industrial mixer at ~6000 rpm for $\approx20\,\mathrm{min}$ we produced polydisperse oil-in-water emulsions, \cite{paredes2013, dekker2018} which remained stable for the duration of the experiments (Sup.~Fig.~1) and beyond.
The fluorophore BODIPY-C12 was synthesized using the method by Lindsey and Wagner \cite{wagner1996boron}.
Prior to the preparation of the emulsions, a stock solution of BODIPY-C12 in ethanol was diluted with the continuous phase ($1:100$) to obtain a dye concentration of $\approx1\,\upmu \mathrm{M}$.
We anticipate the presence of ethanol not to impact the experiments considering its strong dilution.   
BODIPY-C12 dissolves in micellar solution, and in oil, but is poorly soluble in water (Sup.~Fig.~2).
The chemical structures of the dye molecule, and surfactants are drawn in Fig.~\ref{fig:structures}.
\subsection{Fluorescence Lifetime Imaging Microscopy}
All fluorescence lifetime imaging microscopy (FLIM) measurements were carried out using a Leica TCS SP8 HyD confocal fluorescence lifetime microscope.
As an excitation source we used a $470\,\mathrm{nm}$ pulsed laser at $40\,\mathrm{MHz}$ and the emitted light was detected in the range of $500\,\mathrm{nm}-700\,\mathrm{nm}$ by a Hyd detector.
For all experiments we used a 100x magnification in-oil objective with a numerical aperture of $1.25$.
Images were acquired at a scan speed of $100\,\mathrm{Hz}$ and accumulated $4$ times each, in the course of several hours after sample preparation, and for up to six consecutive days after.
From these images we extracted both the fluorescence intensity, which is proportional to the concentration of the dye $c$ (Sup.~Fig.~3), and the fluorescence lifetime, which provides information on the local environments the dye molecules are exposed to \cite{kuimova2012mapping,wu2013,hosny2013,lopez2014molecular, kang2020}.
For the analysis of the fluorescence intensity of the microscopy images we used the image processing package Fiji \cite{schindelin2012} along with the collection of plugins MorpholibJ \cite{legland2016} and the ellipse splitting plugin \cite{wagner2017}.
We refer to the supplementary information for a more detailed description of the image analysis.
The fluorescence lifetime was analysed using the Leica Application Suite X.
For each analyzed phase we included at least $10^4$ photon counts.
The time-resolved fluorescence was fitted using a model based on n-exponential reconvolution.
For the fit of multi-exponential decays we used the amplitude-weighted average lifetime, $\langle \tau \rangle=\frac{\sum_{i}A_{i} \tau_i }{\sum_{i}A_{i}}$.
We considered fits acceptable for $\chi^2<1.5$.

\subsection{Simulation Model and Methods}
To reach the length- and time scales associated with the diffusion of the dye molecules through the oil-water interface and capture experimental trends in a qualitative manner, we chose to model the systems using a coarse-grained (CG) description. These calculations using the atomistic resolution of the compounds would be computationally unfeasible. Additionally, to the best of our knowledge, there is no full-atomistic model for BODIPY-C12 and TX-100 molecules that are compatible with CG models of water, SDS, and silicone oil. Thus, we performed CG molecular dynamics simulations of systems containing a solution of surfactant and dye molecules in a mixture of water and oil. 

As we aimed for a generic model, we deliberately did not take into consideration the shape anisotropy of the solvent particles and modeled them explicitly as spherical beads of unit diameter $\sigma$ and unit mass $m$.
The interaction between solvent particles was modeled via the Lennard-Jones (LJ) potential
\begin{equation}
    U_{\rm LJ}(r) = 4\varepsilon_{ij}\left[\left( \frac{\sigma}{r} \right)^{12} - \left( \frac{\sigma}{r} \right)^{6} \right]
    \label{eq:ULJ},
\end{equation}
where $r$ is the distance between a pair of particles, and $\varepsilon_{ij}$ controls the interaction strength between particles of type $i$ and $j$. For particle pairs of the same type, we used $\varepsilon_{\rm oo} = \varepsilon_{\rm ww} \equiv \varepsilon = k_{\rm B} T$, where $k_{\rm B}$ is the Boltzmann constant and $T$ is the absolute temperature. The cutoff radius of the LJ potential was set to $r_{\rm cut}$ = 3.0 $\sigma$. The interspecies interaction was modeled using the purely repulsive Weeks-Chandler-Andersen (WCA) potential, achieved by truncating  $U_{\rm LJ}(r)$ at its minimum $r_{\rm min}=2^{1/6}\sigma$ and shifting it by $\varepsilon$ \cite{bishop1979molecular}. At the simulation conditions employed, the two fluids are immiscible which leads to the formation of a liquid-liquid interface \cite{roy2016structure}. This strategy was shown to be robust for modeling the immiscibility of two fluids \cite{roy2016structure, diaz2005phase}, including oil-water interfaces \cite{morozova2019surface}.

We approximated both dye and surfactant molecules as chains made up of several beads connected through springs, which were modeled via the finitely extensible nonlinear elastic (FENE) potential combined with the WCA potential. We used the standard Kremer-Grest parameters for the FENE potential to prevent unphysical bond crossing \cite{grest1986molecular}.
In a recent numerical study on the oil-water interfaces decorated by surfactant molecules, a CG mapping for water and SDS molecules was proposed \cite{vu2018oil}. In this representation, five water molecules were lumped together into one bead. Thus, the volume occupied by five water molecules at room temperature, $v_{\rm w}=0.15 \; \rm{nm}^3$, defines the volume of one bead in the simulations. The hydrophobic tail of SDS is equivalent to two water beads in size and is represented by two beads (bead type T). The headgroup (bead type $\rm {H_{s}}$) was modeled by a single bead resulting in three beads per surfactant ($n_{\rm s}$ = 3).  Since SDS and BODIPY-C12 molecules are quite similar with respect to their chemical structure (compare Fig.~\ref{fig:structures}a and b), i.e. both molecules are composed of a hydrophobic tail containing twelve carbon atoms and a headgroup (bead type $\rm {H_{d}}$), we used the same mapping for both BODIPY-C12 and SDS. To investigate how the length of the headgroup of a surfactant molecule influences the diffusion of dye molecules, we additionally introduced a surfactant type that is similar to TX-100 (Fig.~\ref{fig:structures}c). The hydrophobic tail of TX-100 was also modeled by two $\rm T$-type beads, as its length is similar to the hydrophobic tails of SDS or BODIPY-C12 molecules. The head group of TX-100, however, is a polyethylene oxide (PEO) chain. For ten monomers, the Kuhn length for PEO is $\approx0.68\,\mathrm{nm}$, which roughly equals the size of two monomers \cite{liese2017hydration}. Since in our simulations the unit of length $l_u=(6 v_{\rm w}/\pi)^{1/3}\approx0.66 \; \rm nm $, we modeled the headgroup of TX-100 using five $\rm H_{s}$ beads ($n_{\rm s}=7$).
A schematic mapping and the detailed summary of bead types is provided in Fig.~\ref{fig:structures}.

Nonboned interactions between all bead types were also modeled using the $U_{\rm LJ}$ potential introduced above. As the parameter choice is crucial to adequately model the system, we chose the strength of the potential $\varepsilon_{ij}/\varepsilon$ based on the available experimental information to capture the relative strength between the compounds. The resulting values are summarized in Table~\ref{tab:sim_strength}.

\begin{table}
    \centering
	\begin{tabular} {c c c c c c}
	\hline
	 & W & O & $ \rm H_{\rm s}^{\rm short}/\rm H_{\rm s}^{\rm long}$ & $ \rm H_{\rm d}$ & T  \\
	\hline
	W & 1.0 & WCA & 1.9/1.1 & 0.9 & 0.2 \\
	O  &  & 1.0 & 0.2 & 0.9 & 0.9 \\
	$ \rm H_{\rm s}^{\rm short}/\rm H_{\rm s}^{\rm long}$ & & & 1.0 & 3.6/0.225 & 0.2 \\
	$ \rm H_{\rm d}$ &  &  &  & 1.0 & 0.9\\
	T &  &  &  & & 1.0\\
	\hline
	\end{tabular}
	\caption{The values of the interaction strength $\varepsilon_{ij}/\varepsilon$ used in the pair potential $U_{\rm LJ}$ to model nonbonded interactions between $i$ and $j$ particle types.
	}
    \label{tab:sim_strength}
\end{table}

For the intraspecies interaction, we used $\varepsilon_{\rm TT}=\varepsilon_{\rm H_{s} H_{\rm s} } =\varepsilon_{\rm H_{d} H_{d} } = \varepsilon$. In our simplified modeling approach, we use a single interaction strength, $\varepsilon_{min}=0.2\varepsilon$, to capture the immiscibility between some species in the system: $\rm H_{s}/\rm O$, $\rm H_{s}/\rm T$, and $\rm T/\rm W$.
Since the headgroup of the dye molecule is miscible in both types of solvents, we set up the interactions between the headgroup and solvents as $\varepsilon_{\rm H_{d} \rm {O}} = \varepsilon_{\rm H_{d} \rm {W}} = 0.9 \varepsilon$ \cite{zhang2019bodipy}.
For the oil-soluble tail of both the dye and surfactant molecule we also chose $\varepsilon_{\rm H_{\rm d} \rm {T}}/\varepsilon$ to be equal to $0.9$. Experimentally, the surfactant molecules are either found at the oil-water interface screening the interaction between the immiscible fluids or in aqueous solution. To achieve such a scenario in the simulations, we set the interaction strength $\varepsilon_{\rm H_{s} W }$ between the water beads and the headgroup of the surfactants $\rm H_{s}$ to $1.9\,\varepsilon$ and $1.1\,\varepsilon$ for the three- and seven-bead surfactant models, respectively. This choice of interaction parameters results in a similar interfacial coverage by the two surfactant types (Sup.~Fig.~4). 

To qualitatively capture the hydrophilic strength of the surfactant headgroups employed in the experiments, we make use of their HLB values \cite{guo2006calculation}. 
As we represent the SDS headgroup by a single bead (short), while using five beads for the headgroup of TX-100 (long) (Fig.~\ref{fig:structures}b and c), the ratio of the interaction strength is then estimated as $\varepsilon_{\rm H_{\rm s} \rm H_{\rm d}}^{\rm short}/ \varepsilon_{\rm H_{\rm s} \rm H_{\rm d}}^{\rm long} = 38/(12/5) \approx 16 $. Thus, we choose $\varepsilon_{\rm H_s H_d}/\varepsilon$ to be equal to 3.6 and 0.225 for short and long surfactant chains, respectively.

To model the interface, we chose a simulation box elongated along the $z$-direction.
The box size was set to $L_x=L_y=36 \sigma$, and $L_z=3 L_x =108 \sigma$. Due to the periodic boundary conditions there are two interfaces within the simulation box.
In our simulations we determined the position of the interface between the water and oil phases from the maximum of the surfactant concentration profile along the $z$-axis \cite{ren2019molecular}.
The overall particle number density was set to $\rho=0.66 \sigma^{-3}$ \cite{morozova2019surface}. We chose the composition of the solvent mixture as $50:50$, leading to approximately $43000$ particles for each type of liquid. As in the experiments, we assumed that the liquid-liquid interface is saturated with surfactant molecules. We estimate the number of surfactant required as $N_{\rm s}= 2 L_x L_y l_{\rm u}^2 / A_{\rm s}$, where $A_{\rm s} \approx 0.5 \; \rm {nm^2/molecule}$ is the experimentally determined surface area per SDS molecule adsorbed at the interface between silicone oil and water \cite{kanellopoulos1971adsorption}. The resulting values for $N_{\rm s}$ is $2253$ surfactant molecules per simulation for the surfactant model based on three beads. This leads to the surfactant concentration $c_{\rm s}= n_{\rm s} N_{\rm s}/ V_{\rm box}=0.05 \sigma^{-3}$, which is close to the experimental values. For the seven-bead surfactant model, we kept the same concentration resulting in $N_{\rm s}=966$ chains per simulation. We then added $N_{\rm d}=100$ dye molecules into the system to investigate their diffusive behavior. In total, the systems are composed of $92378$ beads.

Starting configurations were generated by randomly placing all oil beads in one half of a simulation box, while placing the remaining beads in the second half. We followed a multi-step equilibration procedure. First, we ran a short simulation for $1 \times 10^6 $ time steps where only the position of the liquid beads was integrated to equilibrate the phase-separated fluids. Next, we achieved a homogeneous distribution of surfactant molecules in a box by simulating a system for $1 \times 10^7$ time steps in which the interaction strength between the T, $\rm H_{s}$ beads and liquids was set to $\varepsilon$. Afterwards, a simulation of $3.5 \times 10^7$ time steps was conducted in which the surfactant and solvent beads interact with the parameters presented in Table \ref{tab:sim_strength}. This simulation results in the formation of a liquid-liquid interface decorated by surfactants. To keep the dye molecules in the aqueous phase during the equilibration stage, we set a purely repulsive interaction (WCA) between the $\rm H_{d}$ and $\rm T$ beads and the oil phase. Finally, we set all interactions according to Table \ref{tab:sim_strength}, and conducted a production run of $2 \times 10^8$ time steps. Three independent runs were performed for each value of the interaction strength $\varepsilon_{\rm H_{\rm s} \rm H_{\rm d}}$. Simulations were conducted in the $NVT$ ensemble at $T=1$ using a Nos{\'e}-Hoover thermostat. The equations of motion were integrated using the velocity Verlet algorithm with a time step of $\Delta t = 0.005 \tau$, where $\tau = \sqrt{m\sigma^2/(k_{\rm B}T)}$ is the intrinsic MD unit of time. All simulations were performed using the HOOMD-blue simulation package (v.2.9.2) \cite{anderson2020hoomd}.

\section{Results and Discussion}

\subsection{Experimental Results}
\begin{figure*}[ht!]
\includegraphics[width=\textwidth]{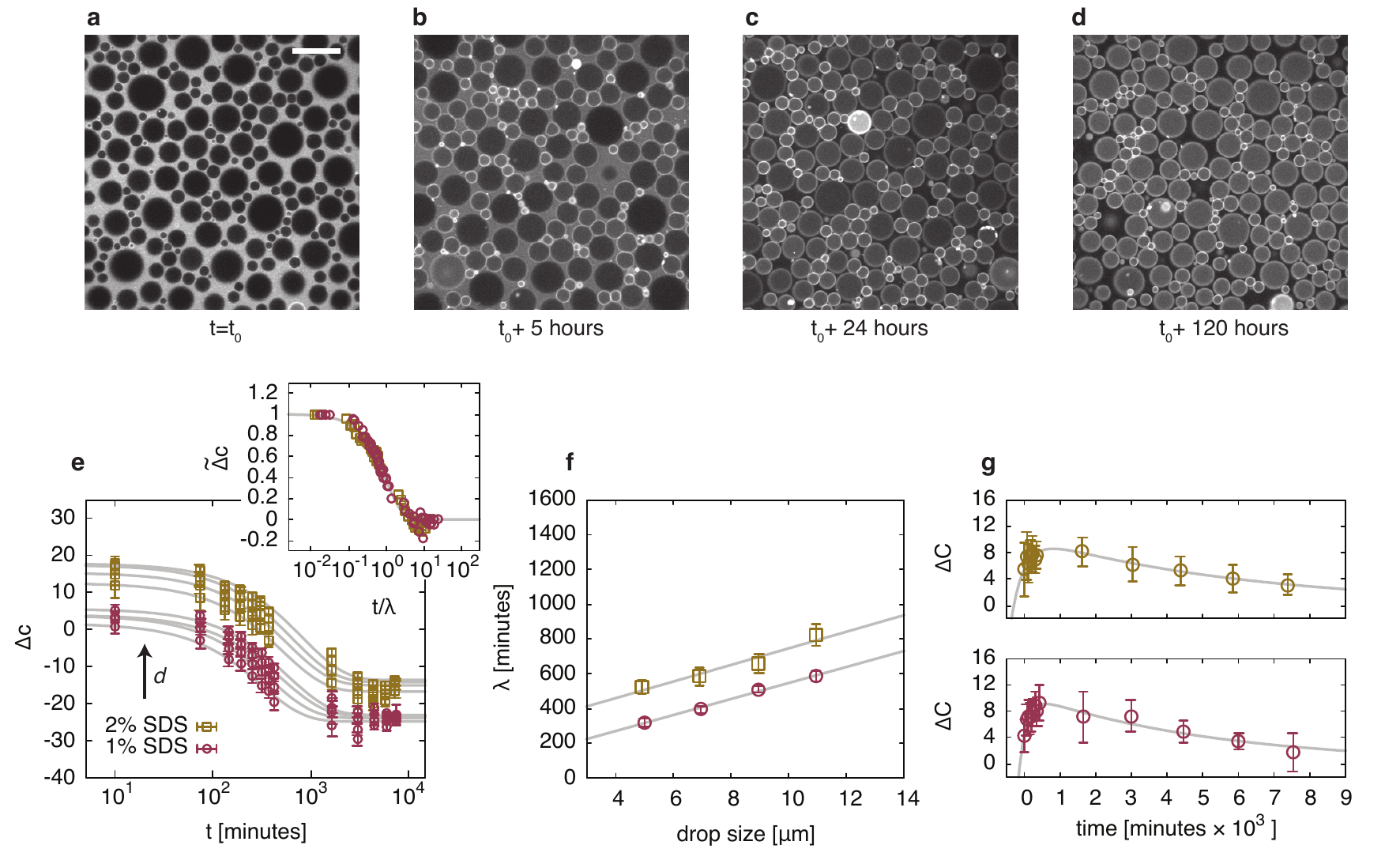}
\caption{Molecular transport in an emulsion. (a-d) Fluorescence microscopy images show the molecular transport of the lipophilic dye BODIPY-C12 from the continuous phase towards the dispersed phase of a model oil-in-water emulsion stabilized with $1\%$ SDS. Once the continuous phase is depleted of BODIPY-C12 (after $\approx24$ hours), the fluorophore is also transported between oil droplets. The whole exchange takes several days. Scale bar is $20\,\upmu \mathrm{m}$. (e) We quantify this exchange using the parameter $\Delta c$, which gives the difference in concentration of dye molecules in the continuous phase and in the oil droplets: $c_\mathrm{w}-c_\mathrm{o}$. For different oil droplet sizes, $d$ (average drop sizes: $\langle d \rangle \approx[5,7,9,11]\,\upmu \mathrm{m}$), $\Delta c$ decays exponentially and is fitted using Eq.~(\ref{eqn:eq1}). Normalizing $\Delta c$ by its initial and final value, and rescaling this normalized concentration difference using a single timescale $\lambda$, collapses all data (inset). (f) The timescale $\lambda$ increases linearly with droplet size with a slope independent of surfactant concentration. (g) To quantify the exchange between the oil droplets we use the parameter $\Delta \mathrm{C}$, which gives the difference in concentration of dye molecules in oil droplets of different sizes: $\Delta c(\langle d \rangle = 5\upmu \mathrm{m})-\Delta c(\langle{d}\rangle = 11\upmu \mathrm{m})$. The dynamics of $\Delta \mathrm{C}$ as a function of time reveal two timescales for both emulsions stabilised with $2\%$ (top) and $1\%$ SDS (bottom).
Initially, $\Delta \mathrm{C}$ grows as the smaller droplets fill up faster (inset). Then, as the continuous phase is depleted of dye molecules, BODIPY-C12 is being transported from concentrated small droplets towards less concentrated large droplets. This process happens at a timescale of days.
The fit is the exponential model presented in Eq.~(\ref{eqn:eq2}).}
\label{fig:Fig2}
\end{figure*}

In the case of SDS stabilized emulsions, at $t_{0}$, which for all samples is $\approx10$ minutes after emulsion preparation, we detect the fluorophore in the continuous phase only (Fig.~\ref{fig:Fig2}a), where it is loaded onto micelles (Sup.~Fig.~2,~5).
BODIPY-C12 then diffuses from the initially swollen micelles into the oil droplets, which happens gradually over the course of $~24$ hours (Fig.~\ref{fig:Fig2}b, c), with a rate that is dependent on the size of the oil droplets; the smaller droplets ``fill up'' faster.
In addition, we observe highly fluorescent regions at the periphery of the droplets, which suggests aggregation of the dye molecules at the oil-water interface \cite{osakai2007potential,zhou2018fluorescence,xiong2020}.
After the continuous phase is depleted of BODIPY-C12 (Sup.~Fig.~6),  we find the dye diffusing from the brighter small droplets towards the empty larger ones until the fluorescence intensity is uniform among the oil phase (Fig.~\ref{fig:Fig2}d). This molecular transport is not limited to an emulsified system, and also occurs in bulk (Sup.~Fig.~7).

We characterize the exchange of dye by first analysing the temporal evolution of the concentration difference $\Delta c=c_\mathrm{w}-c_\mathrm{o}$, which is defined as the difference between the concentration of dye molecules in the water phase, $c_\mathrm{w}$, and the concentration of dye molecules dissolved in the silicone oil droplets, $c_\mathrm{o}$.
To obtain the concentration from the fluorescence intensity, we need to know the different extinction coefficients of BODIPY-C12 in micellar solution and in oil.
We therefore first measure the fluorescence intensity of the dye in both neat phases separately and then correct the concentrations accordingly.

To analyse the dependence of the exchange on the oil droplet size, $d$, we binned $\Delta c(d)$ into four intervals (Sup.~Fig.~8) to obtain $\Delta c(t)$ for the average droplet sizes $\langle d \rangle \approx[5,7,9,11]\,\upmu \mathrm{m}$.  
A plot of $\Delta c$  versus time (Fig.~\ref{fig:Fig2}e) reveals exponentials decays that we fit using
\begin{equation}
 \begin{aligned}[b]
    &\Delta c = \Delta c_\mathrm{0} e^{\frac{-(t-t_\mathrm{0})}{\lambda}} + \Delta c_\mathrm{\infty}.
 \end{aligned}
\label{eqn:eq1}
\end{equation}
Rescaling the normalized data $\Delta \tilde{c}=\frac{\Delta c-\Delta c_\mathrm{\infty}} {\Delta c_\mathrm{0}-\Delta c_\mathrm{\infty}} $ between $\Delta c(t_{0})=\Delta c_\mathrm{0}$ and $\Delta c(t_{\infty})=\Delta c_\mathrm{{\infty}}$ (measured at day 6) by the timescale $\lambda$ produces a master curve (Fig.~\ref{fig:Fig2}e, inset).
This exponential relaxation is in agreement with theory based on diffusive transport facilitated by surfactants; the transport is governed by thermodynamics, and dictated by differences within the chemical potential of the solute over which the system equilibrates \cite{skhiri2012,gruner2016}.
The timescale of the transport was previously found to be determined by the permeability $P$ of the micellar phase, the droplet volume $V$ and surface area $S$, and can be expressed as $\lambda=V/(SP)$ \cite{skhiri2012,gruner2016}.
This relation predicts the timescale to scale linearly with droplet size, which our data confirms (Fig.~\ref{fig:Fig2}f), albeit with an offset $\lambda_{0}$ that varies with the surfactant concentration.
Using the slope of $\lambda (\langle d \rangle)$ (fit in Fig.~\ref{fig:Fig2}f) we derive a permeability $P$, defined as the diffusion rate of the dye molecules through the micellar phase, which regardless of surfactant concentration, is on the order of $ 10^{-10} \mathrm{m\ s^{-1}}$.

We first investigate whether the transport is limited by diffusion of the dye through the continuous phase.
In this case, the permeability can be expressed as $P=\frac{KD_\mathrm{m}}{l}$ \cite{zwolinski1949}, with  $K=\frac{c_\mathrm{o}(t_\mathrm{\infty})}{c_\mathrm{w}(t_\mathrm{\infty})}$ the partition coefficient, $D_\mathrm{m}$ the diffusion coefficient of the micelle-dye aggregates, and $l$ the thickness of the interface.
From the fluorescence images at equilibrium we estimate $K\approx 4$ and  $K\approx 2$ for the emulsions stabilized with 2\,\% SDS and 1\,\% SDS, respectively.
In addition we find that the ratio in $\lambda_{0}$ between 2\% and 1\% SDS is given by the ratio between surfactant concentration in the continuous phase, which has been found to scale linearly with the partition coefficient $K$ \cite{gruner2016}.
Micelles formed by SDS were shown to be $\approx 10^{-9}\,\mathrm{m}$ in diameter \cite{bruce2002molecular}, and are thus, according to the Stokes-Einstein equation, expected to diffuse at $D_\mathrm{m}\approx 10^{-10}\,\mathrm{m^2\ s^{-1}}$. Then, estimating the diffusing object to cross an interface of nanometric thickness $l$, we predict permeabilities a factor of $10^{9}$ larger than what we experimentally observe.

We thus conclude that for the molecular transport of BODIPY-C12 in our oil-in-water emulsions stabilized by SDS, crossing the oil-water interface is the rate limiting step, not micellar diffusion through the continuous phase.
Once the continuous phase is depleted of fluorophore we observe exchange of the dye between the droplets (compare Fig.~\ref{fig:Fig2}c and d) \cite{skhiri2012,gruner2016}.
To quantify this we define the concentration difference between the largest and smallest droplets within the samples, i.e. $\Delta C = \Delta c \left( \langle d \rangle \approx 5\upmu \mathrm{m} \right)-\Delta c(\langle d \rangle \approx11\upmu \mathrm{m})$.
Irrespective of surfactant concentration, $\Delta C$ first increases as a consequence of the depleting continuous phase (Fig.~\ref{fig:Fig2}g), in agreement with the data shown in Fig~\ref{fig:Fig2}e.
Subsequently, $\Delta C$ goes through a maximum  $\Delta C_\mathrm{max}$ followed by relaxation to $\Delta C\rightarrow0$.
Using the previous arguments based on diffusion we model this using
\begin{equation}
\Delta C = (\Delta C_\mathrm{0} - \Delta C_\mathrm{max})e^{[\frac{-(t-t_{0})}{\lambda_{1}}]} + \Delta C_\mathrm{max}e^{[\frac{-(t-t_{0})}{\lambda_{2}}]},
\label{eqn:eq2}
\end{equation}
which combines the depletion of the continuous phase with the exchange of the dye molecules between the droplets.
The initial uptake occurs on a timescale of $\lambda_{\mathrm{1}}\approx 400\,\mathrm{min}$ and $200\,\mathrm{min}$ for the emulsions stabilized with $2\%$ and $1\%$ SDS, respectively.
These values are, as expected, similar to the ones measured for the relaxation of $\Delta c$ between the continuous phase and the oil droplets (Fig.~\ref{fig:Fig2}f), i.e. $\lambda$ obtained from $\Delta c$ $\approx$ $\lambda_{\mathrm{1}}$ obtained from $\Delta C$.
The relaxation describing the exchange between droplets happens on much longer timescales of $\lambda_{\mathrm{2}}\approx 6200\,\mathrm{min}$ for $2\%$ SDS and 5400$\mathrm{min}$ for $1\%$ SDS.
We note that the non-overlapping values for $\Delta c(t_{\infty})$ and the fact that $\Delta C>0$ at $t_\mathrm{{\infty}}$ suggest that the smaller droplets still exhibit slightly higher concentrations in dye. This small, but noticeable effect might arise from either experimental errors, or the possibility that the system did not fully reach equilibrium at the last measuring point.
Considering the structural similarities of BODIPY-C12 and SDS with respect to their nonpolar alkyl moieties (compare Fig.~\ref{fig:structures}a and b), it is reasonable to assume that dye-surfactant interactions affect the transport rate.

\begin{figure}
\centering
\includegraphics[width=\columnwidth]{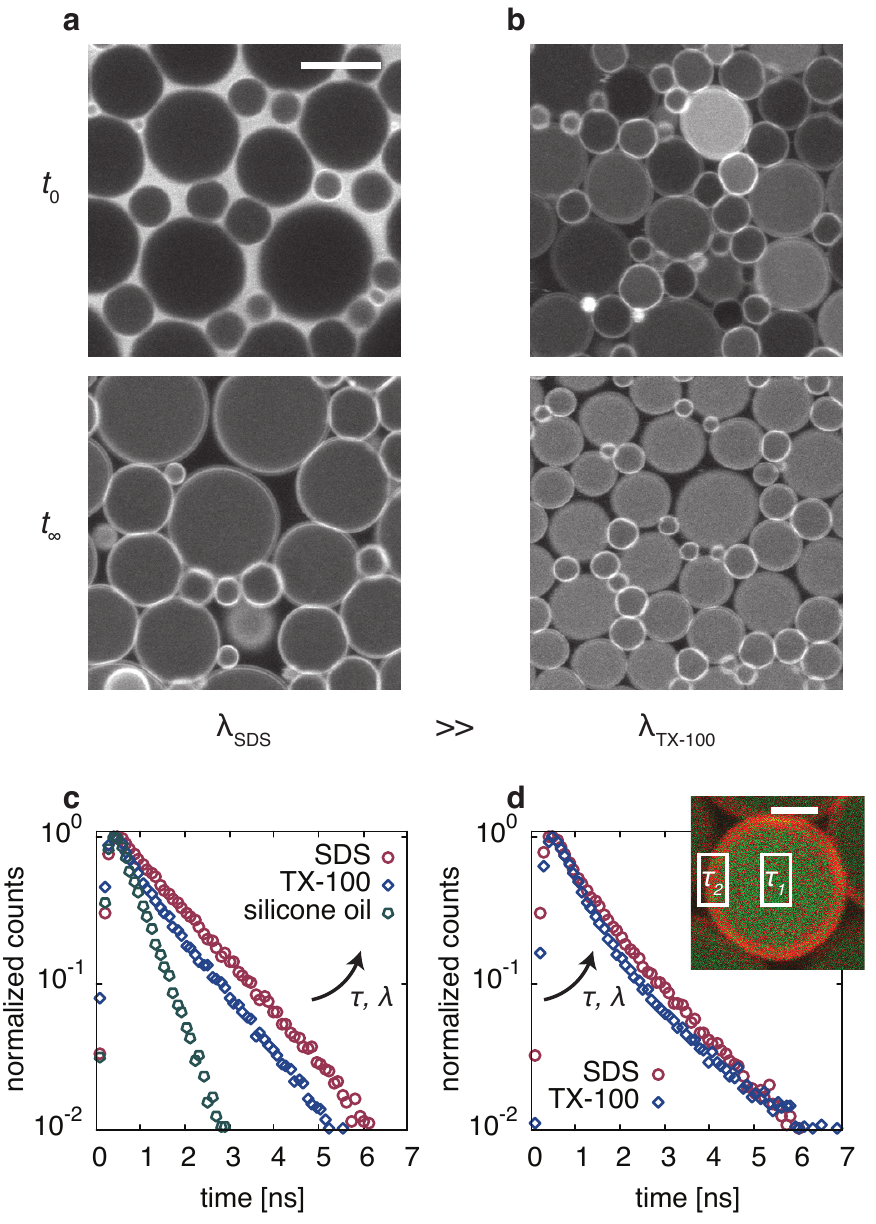}
\caption{Effect of the surfactant on the molecular transport. (a, b) Fluorescence intensity images of BODIPY-C12 in SDS (a) and TX-100 (b) stabilized emulsions recorded after the emulsions were prepared, at $t_\mathrm{{0}}$, and after a couple of days when all oil droplets equilibrated to equal dye concentrations, at $t_\mathrm{{\infty}}$. Scale bar is $10\,\upmu \mathrm{m}$.
The fluorescence decays of BODIPY-C12 recorded in micellar solutions (c) and at the bright oil-water interface (d) show faster decays for TX-100. The extracted lifetime values are summarized in Table~\ref{tab:table2}.
(d, inset) Using the two components found at the oil-water interface we construct a FLIM image, shown here for a single oil droplet stabilized with TX-100. The fast component corresponds to the oil phase (green), while the slow component shows BODIPY-C12 in the surfactant phase (red). Scale bar is $2\,\upmu \mathrm{m}$.}
\label{fig:Fig3}
\end{figure}

The degree to which surfactant molecules are hydrophilic or lipophilic can be estimated using their hydrophilic-lipophilic balance (HLB), which is based on the molecular structure of the emulsifier.
Using the advanced technique proposed by Guo and coworkers \cite{guo2006calculation} we calculate a HLB number of $39.7$ for SDS, reflecting its strong hydrophilicity.
Substituting SDS by the more hydrophobic  TX-100 (compare Fig.~\ref{fig:structures}b and c), leads to an exchange at the time scale of the emulsion preparation itself, since at $t_{0}$, the continuous phase is already depleted of BODIPY-C12 (compare Fig.~\ref{fig:Fig3}a and b). The TX-100 emulsion was also prepared in clear excess of its critical micelle concentration (at $1\, \mathrm{wt}\%$, $\approx70$ $\times$ cmc) \cite{tiller1984hydrogenation}.
However, substitution of SDS with the structurally similar but more hydrophobic SDBS (compare Fig.~\ref{fig:structures}b and d), results in hardly any increase in the transport rate (Sup.~Fig.~9).
This is surprising considering the comparable HLB values of SDBS and TX-100, which we calculate as $10.7$ and $13.7$, respectively.
To rule out any electrostatic effects slowing down the transport (both SDS and SDBS are anionic) we repeated the experiment with $1\, \mathrm{wt}\%$ \ce{NaCl} added to the continuous phase.
In this case the timescale of the molecular transport of BODIPY-C12 increases, which indicates a change in the partition coefficient (Sup.~Fig.~9) \cite{gruner2016}.
From these additional experiments we can conclude that neither the HLB value nor electrostatic effects can explain the increased transport rate of the dye molecule in the case of emulsions stabilized with TX-100.

To get at the core of this, we resort to the analysis of the time-resolved fluorescence of BODIPY-C12 to study its local environments when exposed to both surfactants, in micellar solutions, and at the oil-water interface.
The fluorescence decay of BODIPY-C12 was shown to be monoexponential with a lifetime sensitive to viscosity, but insensitive to polarity \cite{chung2011, kuimova2012mapping, wu2013, lopez2014molecular}.
The relationship between fluorescence lifetime $\tau$ and macroscopic solvent viscosity $\eta$ is given by Förster-Hoffmann's equation $\tau\propto k\eta^x$, where $k$ and $x$ are empirical constants obtained from calibration with solvents of known viscosity (Sup.~Fig.~10) \cite{forster1971viskositatsabhangigkeit}.
We anticipate the local viscosity to vary with surfactant structure given the presence of large gradients in lateral pressure between the headgroups and tails \cite{seddon2009}.
The fluorescence decay curves of BODIPY-C12 in oil, micellar solutions and at the oil-water interface for both SDS and TX-100 are shown in Fig.~\ref{fig:Fig3}c and d, respectively.
The extracted (amplitude-weighted average) lifetimes and corresponding viscosities $\eta_{loc}$, which we determined using F\"orster-Hoffmann's equation, are summarized in Table~\ref{tab:table2}.
The fluorescence of BODIPY-C12 in micellar solutions of SDS decays monoexponentially with a lifetime $\tau$ = 1.28 $\mathrm{ns}$ (Fig.~\ref{fig:Fig3}c), indicating that the dye molecules are exposed to a single environment.
\begin{table}[t]
\caption{\label{tab:table2}Time-resolved fluorescence parameters. The local viscosities $\eta_{loc}$ are inferred from the amplitude-weighted average lifetimes $\langle\tau\rangle$, using a calibration measurement of ethanol-glycerol mixtures (Sup.~Fig.~10).}

\begin{tabular}{ c c c c c c c }
 \hline
  & $\tau_{\mathrm{1}}$\,[ns] & $\tau_{\mathrm{2}}$\,[ns] & $\frac{A_{1}}{A_{2}}$&  $ \langle\tau\rangle\,[ns]$ & $\eta_{loc}$\,[mPas] & $\chi^2$ \\
 \hline
 $\mathrm{SDS_{aq.}}$& 1.28 &   &  & 1.28 & $55$ & 1.154  \\
 $\mathrm{TX-100_{aq.}}$ & 0.54 &  1.37 & 1.2 & 0.92& $25$ & 0.976  \\ 
 $\mathrm{oil}$ & 0.52 &   &  & 0.52 & $6$ & 1.367  \\
 $\mathrm{SDS_{o-w}}$ & 0.66 &  1.56 & 2.1 & 0.96 & $28$ & 1.044  \\ 
 $\mathrm{TX-100_{o-w}}$ & 0.55 &  1.95 & 6.1 & 0.75& $15$ & 1.029  \\ 
 \hline
\end{tabular}
\end{table}

In this environment, according to our calibration, BODIPY-C12 molecules experience a viscosity of $55\,\mathrm{mPa \ s}$, which implies a significantly lower mobility of the dye than the one measured in bulk solution at low viscosities (Sup.~Fig.~10).
The fluorescence decay of BODIPY-C12 in TX-100 micelles, however, is biexponential, which indicates that the dye molecules probe a second local environment.
The amplitude-weighted average lifetime of BODIPY-C12 within TX-100 micelles relates to a lower viscosity of $25\,\mathrm{mPas}$.
Strikingly, the viscosity experienced by BODIPY-C12 in the oil phase appears to be decoupled from the macroscopic viscosity, i.e. the local viscosity $\eta_{loc}$ is orders of magnitude lower than the one reported from conventional rheometry \cite{vu2016tuning, polita2020effect, bittermann2021disentangling}.
This might also explain the lower values for $\langle\tau\rangle$ and $\eta_{loc}$ at the interface, where the presence of oil inevitably affects the lifetime.
Indeed, fitting both components to construct a FLIM image (Fig.~\ref{fig:Fig3}d, inset) reveals that the fast component belongs to the oil phase. 
Regardless of whether BODIPY-C12 resides in the micellar phase or at the oil-water interface, samples prepared with TX-100 instead of SDS show a local viscosity a factor of $2$ lower.

\begin{figure}[t!]
\centering
\includegraphics[width=\columnwidth]{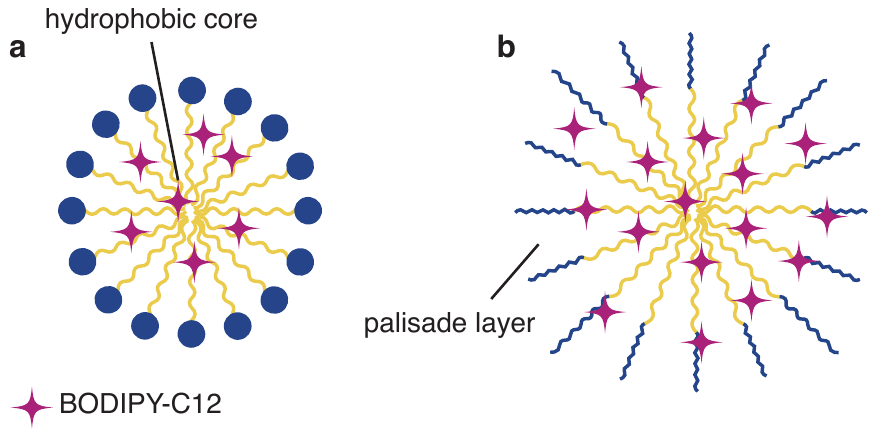}
\caption{Proposed local environments of BODIPY-C12 in micelles. As suggested by the analysis of the fluorescence decay of micellar solutions, BODIPY-C12 is more localized in SDS micelles (a). 
In this case, its exchange between micelles and micelles and oil droplets is limited.
In the larger TX-100 micelles, however, BODIPY-C12 is less localized, and also populates the outer palisade layer (b), which facilitates a fast exchange.}
\label{fig:Fig4}
\end{figure}

We tentatively interpret these results as follows; given the structural similarities between the surfactant molecules and BODIPY-C12 we hypothesize that the dye molecules intercalate into the micelles. In the case of SDS the dye molecules are subject to a higher degree of molecular crowding and are thus more stable within the macromolecular assembly. Within TX-100 micelles however, BODIPY-C12 molecules are less localized and more mobile, as suggested by the biexponential decay and significantly lower lifetime. One possible explanation would be that the dye molecules in the TX-100 case are populating both the hydrophobic core of the micelles and their outer palisade layer \cite{kumbhakar2004,kumbhakar2005}, which is composed of a polyethylene oxide chain.
Hence, for SDS, the exchange to neighboring micelles \cite{rharbi2001, rharbi2002} and oil droplets is more restricted, compared to the TX-100 case.
These considerations are illustrated in Fig.~\ref{fig:Fig4}.
From a structural perspective, the differences in mobility could also be related to differences in the surface density between SDS and TX-100 micelles, where the much less dense surface of TX-100 micelles could facilitate the fast exchange. Additional structural information on the micelle-dye aggregates could for instance be inferred from X-ray studies.
From our experiments we thus conclude that neither the HLB nor electrostatic effects determine the transport rate of the lipophilic dye within this emulsion, but rather the mobility of the dye molecules within the micelle-dye assemblies.
We further substantiate these claims by performing molecular dynamics simulations. 

\subsection{Simulation Results}

\begin{figure*}[ht!]
\centering
    \includegraphics[width=0.8\textwidth]{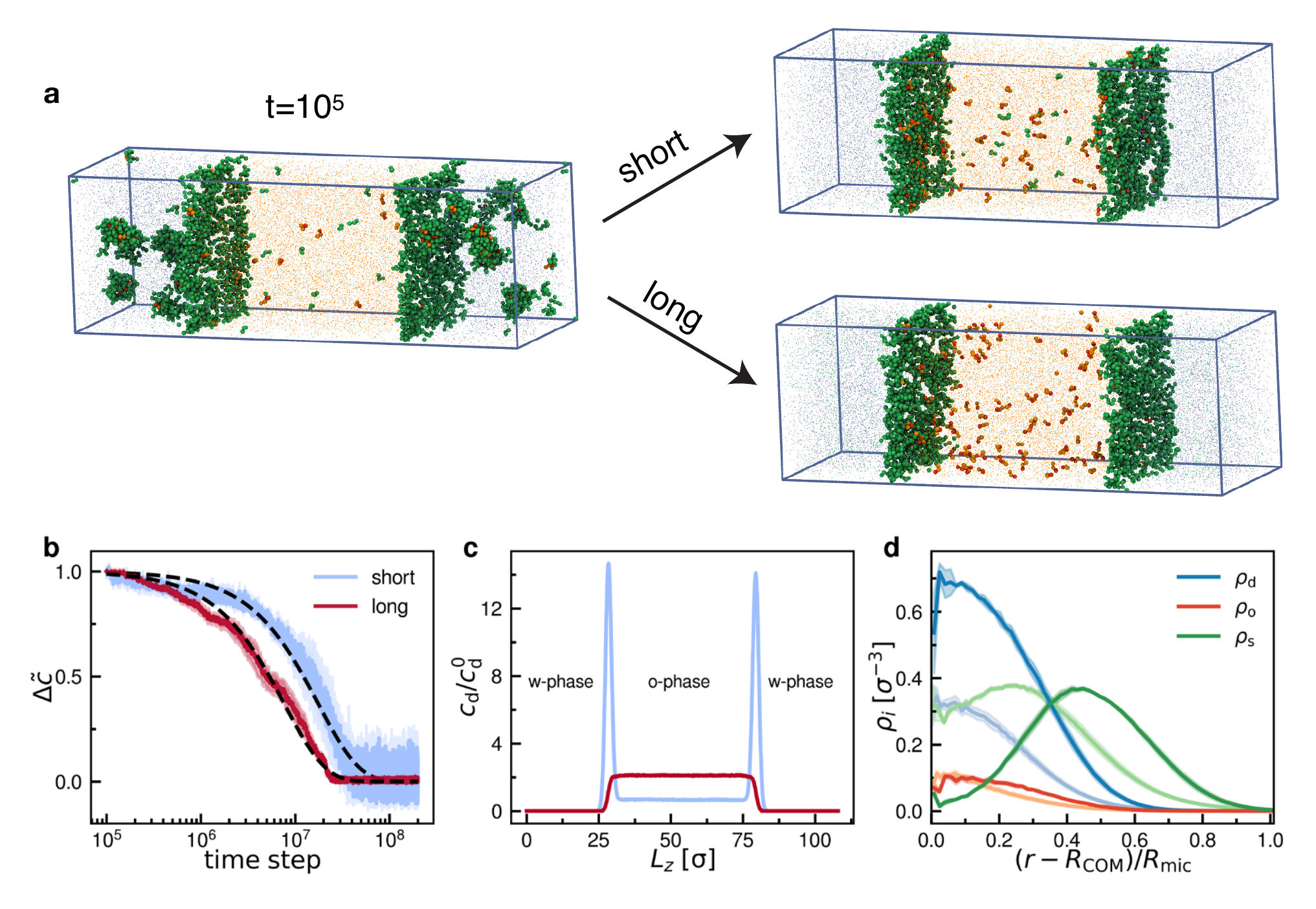}
    \caption{Simulations of the dye diffusion through a oil-water interface decorated by short ($n_{\rm s}=3$) and long ($n_{\rm s}=7$) surfactants. (a) Representative simulation snapshots that correspond to the formation of micelles at $t=10^5$ time steps and to the end of a simulation. The snapshots have been rendered using Visual Molecular Dynamics \cite{humphrey1996vmd}. Dye molecules are colored red ($H_{\rm d}$) and orange (T), surfactant chains are colored green, and oil and water beads are colored orange and blue, respectively. (b) Temporal evolution of the normalized concentration difference $\Delta \tilde {c}$ of dye molecules present in both water and oil phases. The data is fit using $e^{-t/\lambda}$ (dashed lines). (c) The concentration profiles $c_{\rm d}/c_{\rm d}^0$ of beads belonging to dye molecules along the $z$-axis normalized by the bulk values. (d) The number density of beads belonging to a dye ($\rho_{\rm d}$), oil ($\rho_{\rm o}$), or surfactant molecule ($\rho_{\rm s}$) with respect to the radial distance from the center of mass of a micelle, $R_{\rm COM}$, normalized by its size $R_{\rm mic}$. The light and dark curves correspond to systems with short and long surfactant chains, respectively. Shaded areas represent the error bars, calculated as the standard error of the mean from three independent realizations.}%
    \label{fig:comp_short_long}
\end{figure*}

In our simulations, we observe the rapid formation of micellar aggregates (composed of dye, oil, and surfactant molecules) in the water phase during the initial $10^5$ time steps.
We characterize the formation of these aggregates using the density-based clustering algorithm DBSCAN \cite{scikit-learn}. For this clustering we consider O (oil), T (hydrophobic tail), $\rm H_{\rm s}$ (hydrophilic headgroup of the surfactants) or $\rm H_{\rm d}$ (hydrophilic headgroup of the dye) beads (Fig.~\ref{fig:structures}) located in the water phase, and whose $z$-coordinates are at least $3 \sigma$ away from the position of the interface. This ensures that the micelles do not interact with the interface when their composition is analyzed. Dye molecules, not participating in the formation of micelles, diffuse freely through the interface. In the experiments, at $t_{0}$, the dye molecules are already partitioned into micelles in the aqueous phase. Hence, we excluded the first $10^5$ time steps from the analysis of the diffusion of the dye.

Fig.~\ref{fig:comp_short_long}a shows representative snapshots of the systems investigated, at the time of the micelle formation and at the end of a simulation.
Fig.~\ref{fig:comp_short_long}b shows a plot of the temporal evolution of the normalized concentration difference $\Delta \tilde {c}$ of the dye molecules present in the water and oil phase using the same definition for $\Delta \tilde {c}(t)$ as employed in the experiments. 

The simulation runs between $\Delta c_0(t=10^5)$ and $\Delta c_{\infty}(t=2\times 10^8)$ and recovers the exponential decay [Eq.~(\ref{eqn:eq1})] that we also observed in the experiments (Fig.~\ref{fig:Fig2}d, e). The results suggest a slower exchange for the diffusion of solutes in systems where the interface is made up of short surfactant chains than in the ones with long surfactant chains. The characteristic time scale obtained from fitting reveals that for the short surfactant chains $\lambda^{\rm short}=1.8 \times 10^7$ time steps is more than two times larger than the value for long surfactant chains $\lambda^{\rm long}=7.7 \times 10^6$ time steps. These findings are in good qualitative agreement with the experimental results (Fig.~\ref{fig:Fig3}a, b). Thus, our computational results suggest that the energetic penalty to cross the interface decorated by the long surfactant is lower than for the interface decorated by short surfactant as supported by the temporal evolution of the nonbonded potential energy in the system (Sup.~Fig.~11). For a more quantitative description, calculations of the transfer free energy of a dye molecule from an aqueous to an oil phase through the interface decorated by two surfactant types could be performed.

Next, we compute the concentration profile of beads belonging to dye molecules along the $z$-axis (Fig.~\ref{fig:comp_short_long}c). For systems composed of surfactants with long headgroups, dye molecules are homogeneously distributed inside the oil phase. In contrast, dye molecules are more prone to stick to the interface when the headgroup of the surfactant is small.
We note that due to a rather small number ($N_{\rm d}=100$) of dye molecules simulated and on the timescales accessible in the simulations, the difference in the concentration profiles between these systems is amplified. Qualitatively, it resembles the experiments at the early stage, where highly fluorescent regions at the periphery of the
droplets were observed (Fig.~\ref{fig:Fig2}b).
Experimentally, such aggregation was shown to occur for a variety of molecules, independent of charge, and could be attributed to the presence of a gradient in the electric field at the oil-water interface \cite{xiong2020}.
For future studies, it could be interesting to elucidate the mechanism behind the aggregation in more detail, for instance by performing free energy calculations as described above or by performing neutron or X-ray scattering experiments.

Finally, we study the composition of the micelles by calculating the number density profiles of each component. For both simulated systems, the micelles are predominantly made up of dye and surfactant beads, with a small number of oil beads present at the core (Fig.~\ref{fig:comp_short_long}d).  
We find a comparable amount of surfactant beads in both simulated systems, and that the density of dye beads is higher for systems containing weakly interacting headgroups.
Consequently, clusters containing surfactant molecules with long headgroups are less stable against dissolution; this results in a faster diffusion of dye molecules into the oil phase, which is in line with our experimental results, showing a faster diffusion of the dye molecule for emulsions stabilized with TX-100 (long headgroups) than for the ones stabilized with SDS (short headgroups).

\section{Conclusion}

To conclude, we observe experimentally that the molecular transport of a lipophilic dye molecule in a model oil-in-water emulsion is a two-step process that depends on droplet size, surfactant concentration, and surfactant type, for which we suggest simple models. For some surfactants, this transport can be limited by diffusion through the interface, a phenomenon that we find to be independent of both the hydrophilic-lipophilic balance of the surfactant molecules and the presence of electrostatic effects. Instead, the analysis of the time-resolved fluorescence of the molecular rotor suggests that the mobility of the dye molecules within the micelles plays a role; micelle-dye assemblies consisting of surfactant molecules with smaller headgroups stabilize the dye molecules against dissolution into the oil phase. Our findings are supported by molecular dynamics simulations, which recover the behavior of dye depletion from the continuous phase. They show that there is indeed a strong dependence of this molecular transport on the molecular size of the surfactant molecules stabilizing the oil-water interface.
We believe these results can be valuable for designing any application in which emulsions are being used as compartments, particularly for drug delivery \cite{lu2010emulsions}.

\begin{acknowledgments} 

M.~R.~B. acknowledges Nico Schramma for fruitful discussions on image processing.
We thank Hans Sanders for synthesizing the dye molecule.
This work was performed using HPC resources (GPU-accelerated partitions of the Jean Zay supercomputer) from GENCI–IDRIS (Grant 2021 - A0100712464).

\end{acknowledgments}
\newpage

\renewcommand{\figurename}{SUP. FIG.}
\setcounter{figure}{0}

\section*{Appendix: Supplementary Information}
\subsection*{Emulsion Stability}
The droplet sizes of the emulsions are shown in Sup.~Fig.~1 and remained constant over time. 
\begin{figure}[h!]
\centering
\includegraphics[width=\columnwidth]{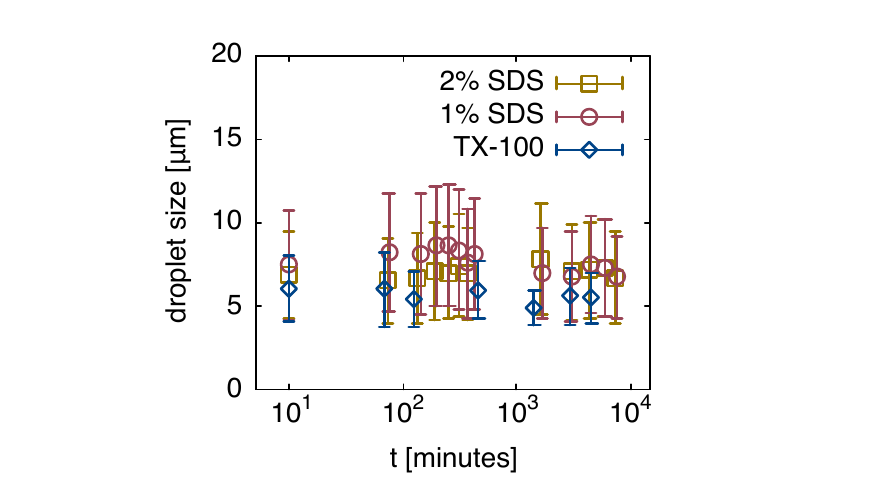}
\caption{Droplet size as a function of time for different surfactant types and concentrations.} 
\label{fig:FigS1}
\end{figure}
\subsection*{BODIPY-C12 Solubility}
Fluorescence microscopy intensity images (Sup.~Fig.~2) show that for micellar solutions below the cmc, BODIPY-C12 forms aggregates and is poorly soluble. 
Above the cmc, however, the intensity is homogeneous. This suggests that BODIPY-C12 is solubilized by micelles. 
\begin{figure}[h!]
\centering
\includegraphics[width=0.8\columnwidth]{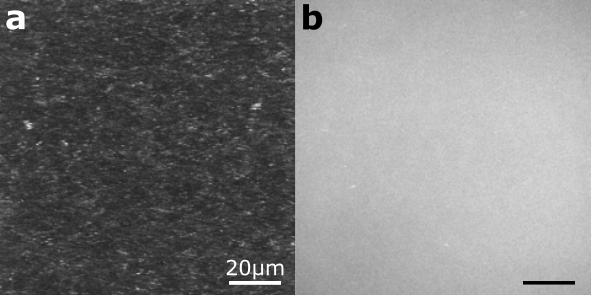}
\caption{Fluorescence intensity images of BODIPY in SDS solutions below the cmc ($0.1\%$, a) and above the cmc ($1\%$, b).}
\label{fig:FigS2}
\end{figure}
\subsection*{BODIPY-C12 Intensity-Concentration Dependence}
To verify the dependency of the intensity of BODIPY-C12 on its concentration we carried out a reference measurement in $1\%$ SDS solution, in which we varied the concentration of the fluorophore. Sup.~Fig.~3 shows an approximately linear relationship.
\begin{figure}[h!]
\centering
\includegraphics[width=0.8\columnwidth]{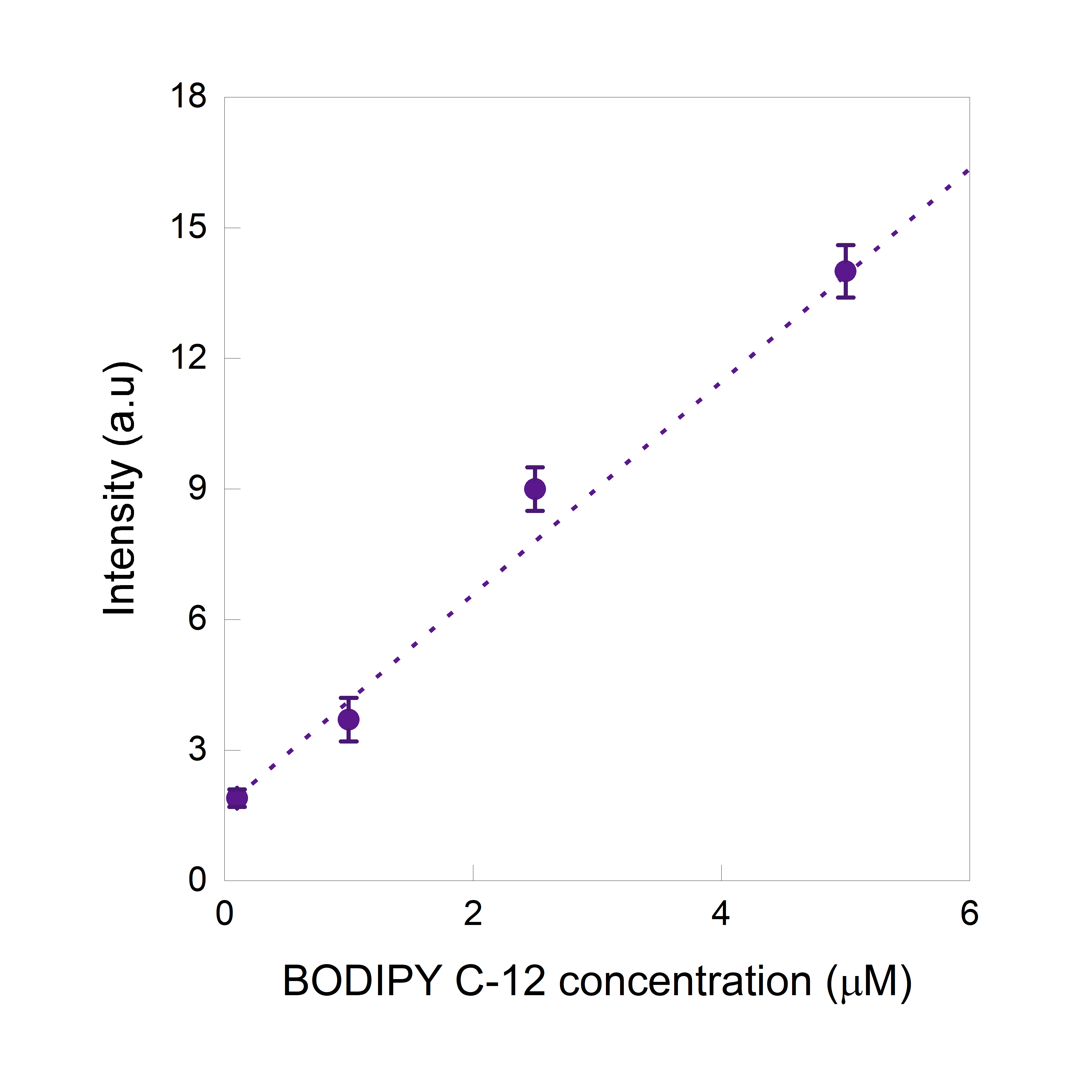}
\caption{Fluorescence intensity plotted versus fluorophore concentration shown for a solution containing $1\%$ SDS. The fit is linear.}
\label{fig:Fig_concentration_dependence}
\end{figure}
\subsection*{Concentration Profile of Surfactant Beads Along the $z$-Axis}
In Sup.~Fig.~4 we plot the concentration profile of surfactant beads along the $z$-axis for two surfactant architectures, i.e. chains composed of three beads (short) and seven (long) beads.\\
\begin{figure}[h!]
\centering
\includegraphics[width=0.8\columnwidth]{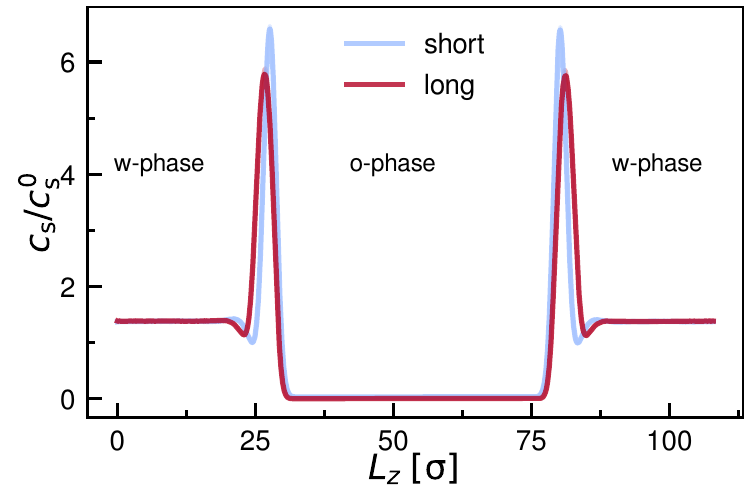}%
\caption{Concentration profile $c_{\rm s}/c_{\rm s}^0$ of beads belonging to surfactant along the $z$-axis normalized by its bulk value for short ($n_{\rm s}=3$) and long ($n_{\rm s}=7$) chains investigated. The interaction strength $\varepsilon_{\rm H_s H_d}/\varepsilon$ equals to 3.6 and 0.225 for a short and a long surfactant chain, respectively.
}
\label{fig:surf_long_short}
\end{figure}
\vspace{-0.6cm}
\subsection*{DLS Measurements of Micelles with/and without the Addition of BODIPY-C12}
The effect of the addition of BODIPY-C12 to the micellar size of SDS and TX-100 is demonstrated by DLS measurements (Sup.~Fig.~5). In both cases, the fluorophore causes the micelles to become bigger, which indicates that the micelles are swollen with BODIPY-C12.
\begin{figure}[h!]
\centering
\includegraphics[width=0.8\columnwidth]{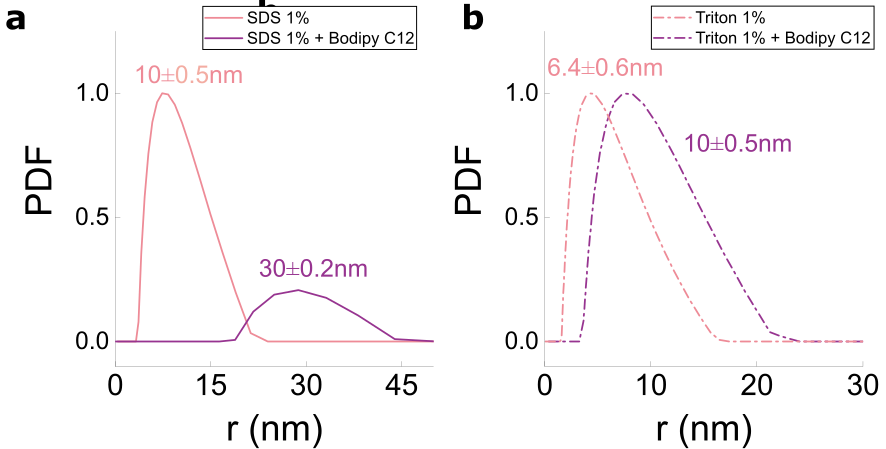}
\caption{DLS measurements of micellar solutions of SDS and TX-100. The addition of BODIPY-C12 shifts the size distributions to bigger radii.}
\label{fig:FigS3}
\end{figure}
\subsection*{Depletion of the Continuous Phase}
Sup.~Fig.~6 shows the decrease of fluorescence intensity of BODIPY-C12 in the continuous phase of SDS stabilized emulsions. Given that SDS micelles are insoluble in oil, we surmise the decrease in intensity to originate from a decrease in dye concentration. From day 2 onwards, the fluorophore concentration in the micellar phase remains constant.
\begin{figure}[h!]
\centering
\includegraphics[width=\columnwidth]{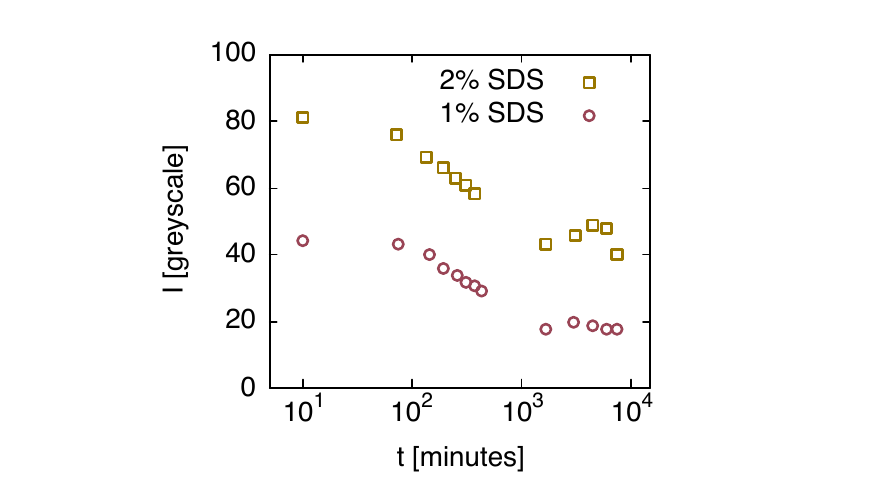}
\caption{Raw data of the fluorescence intensity of BODIPY C-12 in the continuous phase. The concentration of BODIPY-C12 in the continuous phase decreases only on the first day of the measurement.} 
\label{fig:FigS5}
\end{figure}
\subsection*{Molecular Transport in a Non-emulsified System}
To investigate whether the molecular transport of BODIPY-C12 is limited to  emulsified systems, we carried out additional experiments, in which we measured the transport in non-emulsified systems.
For this we squeezed 0.2 mL of micellar phase (containing 1\% and  2\% SDS, and $\approx1\,\upmu \mathrm{M}$ BODIPY-C12) surrounded by 0.8 mL of oil between two glass slides separated by $300\,\upmu \mathrm{m}$ (using spacers) and measured the fluorescence intensity and lifetime analogously to the emulsified system (Sup.~Fig.~7, shown here for 1\% SDS). We find that, after one week ($t_{\infty}$ of the emulsified system) the concentration of dye in the system containing 1\% SDS (Intensity, $I=5.41\pm 0.1$) exceeds the concentration of dye in the system containing 2\% SDS ($I=2.52\pm 0.2$) by a factor $\approx2$, similar to the ratio of the partition coefficients inferred from the emulsified systems. The measured lifetimes in the oil phase were $\tau$ = $0.546\,\pm0.01 \,\mathrm{ns}$ and $\tau$ = $0.562\,\pm0.02\,\mathrm{ns}$ for the samples prepared with 1\% SDS and 2\% SDS, respectively. These values are expected and in line with the lifetimes measured in the oil droplets. However, this bulk system is not at equilibrium yet, indicated by (i) the presence of a diffusion front propagating from the water phase, and (ii) the water phase being brighter than the oil phase. We thus surmise that the transport can also be observed in bulk, albeit on a much slower rate.
\begin{figure}[h]
\centering
\includegraphics[width=\columnwidth]{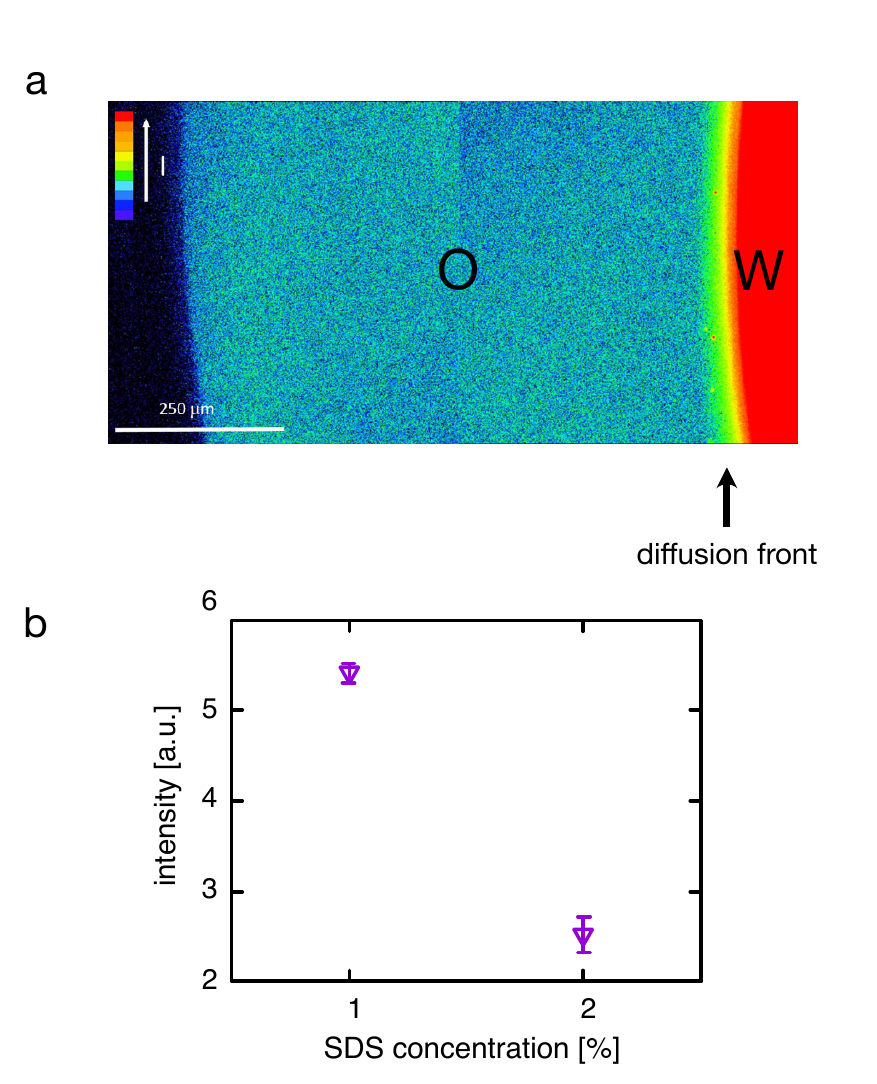}
\caption{(a) Diffusion of BODIPY-C12 from micelles (W, 1\% SDS) into oil (O) in the non-emulsified equivalent of our experiment. The fluorescence intensity image was recorded after one week. The transport happens much slower, as highlighted by the presence of a diffusion front emanating from the intense water phase. (b) The intensities of the oil phases differ by a factor of $\approx 2$.} 
\label{fig:bulk_equivalent}
\end{figure}
\subsection*{Oil Droplet Size Dependence of the Molecular Transport}
A plot of $\Delta c$ versus droplet size for different time steps and concentrations of SDS (Sup.~Fig.~8a, b) reveals that $\Delta c$ increases with droplet size, $d$, but decreases in time.
Interestingly, $\Delta c$ also increases with surfactant concentration. For further data analysis we binned the data into four intervals to fix the droplet size. To mitigate the potential effect of oil droplet size on the intensity signal, as caused by the non-transparency of the emulsions, we kept the range of investigated droplet sizes small, from $4-12\,\upmu m$. 
\begin{figure}[h]
\centering
\includegraphics[width=\columnwidth]{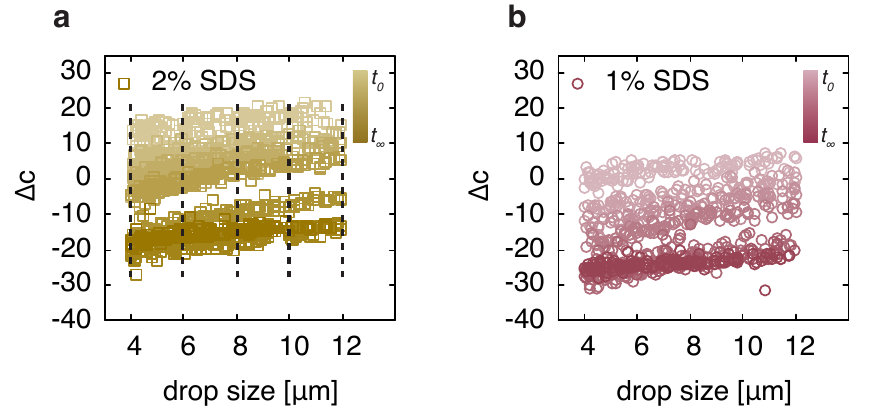}
\caption{(a, b) A plot of $\Delta c$ versus oil droplet size for different time steps (color gradient) and concentration of surfactant. The data is binned into four intervals to fix the drop size (dashed lines). } 
\label{fig:binned}
\end{figure}
\subsection*{Influence of HLB and Salt}
To study the influence of a lower hydrophilic-lipophilic balance (HLB) and salt on the dynamics of the transport process we repeated the experiments with $1\, \mathrm{wt}\%$ SDBS ($\approx76$ $\times$ cmc) \cite{shah2011investigation}, and $1\, \mathrm{wt}\%$ SDS + $1\, \mathrm{wt}\%$ \ce{NaCl}, respectively. Both cases do not significantly speed up the molecular transport, as shown in Sup.~Fig.~9.
\begin{figure}[h!]
\centering
\includegraphics[width=\columnwidth]{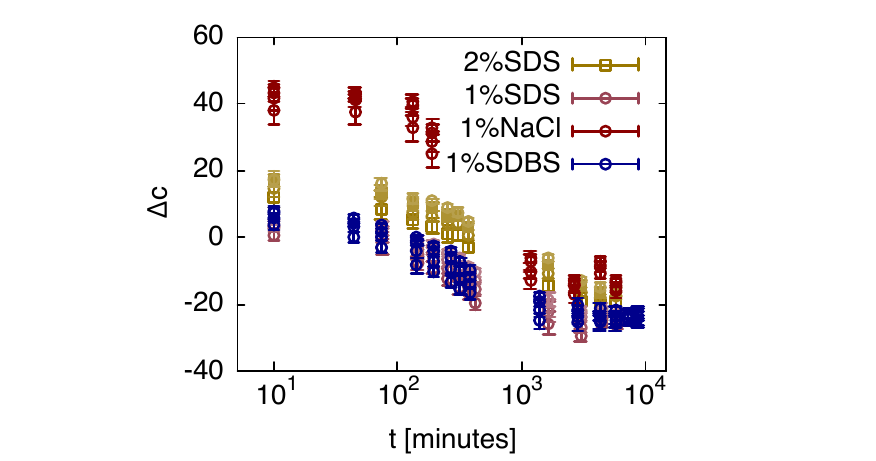}
\caption{Dynamics of the molecular transport of BODIPY-C12 from the continuous phase into oil
droplets. Lowering the HLB value or adding salt to SDS does not slow down the timescale related to the transport process.} 
\label{fig:FigS7}
\end{figure}
\subsection*{Calibration of BODIPY-C12}
According to Förster-Hoffmann's equation the fluorescence lifetime $\tau$ of a molecular rotor scales with solvent viscosity $\eta$ as $\tau\propto k\eta^x$. To calibrate the molecular rotor we prepared ethanol-glycerol solutions with BODIPY-C12 at $\approx 1\,\upmu \mathrm{M}$ and measured the fluorescence lifetime of the solutions.
The solvent viscosities were measured using an Anton Paar MCR 302 rheometer with a cone-plate of $50\,\mathrm{mm}$ diameter at an angle of $1\,^{\circ}$.
From the calibration curve (Sup.~Fig.~10) we obtained $k=0.24\pm0.07$ and $x=0.42\pm0.05$. 
\begin{figure}[h]
\centering
\includegraphics[width=\columnwidth]{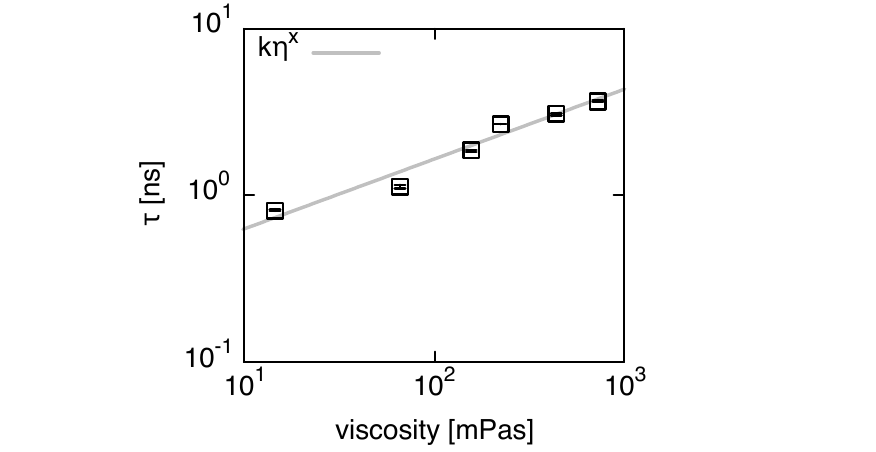}
\caption{Calibration curve of BODIPY-C12 in ethanol-glycerol mixtures. The fluorescence lifetime of BODIPY-C12 increases with solvent viscosity. } 
\label{fig:FigS9}
\end{figure}
\subsection*{Nonbonded Potential Energy}
\begin{figure}[h]
\includegraphics[width=\columnwidth]{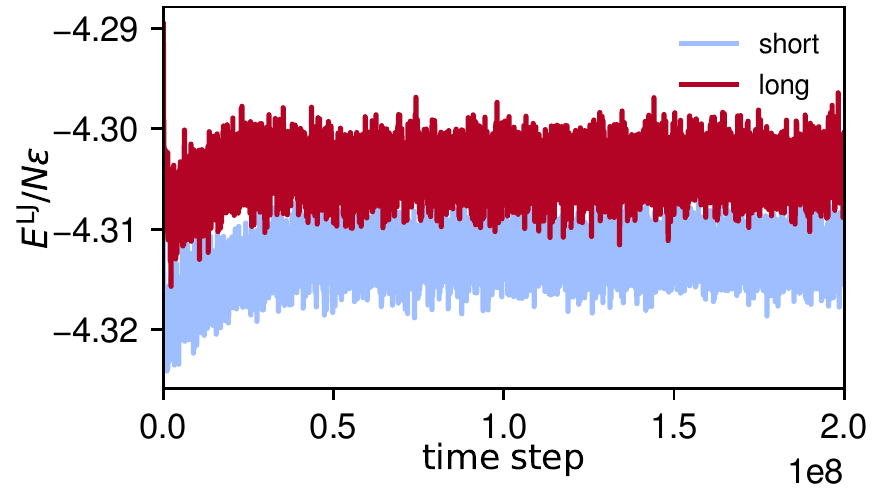}
\caption{Nonbonded potential energy per particle, $U^{\rm LJ}/N \varepsilon$, as a function of the simulation time step for systems containing short and long surfactant chains.} 
\label{fig:FigS8}
\end{figure}
\subsection*{Image Analysis}
Image analysis was carried out using Fiji \cite{schindelin2012} together with MorpholibJ \cite{legland2016} and the ellipse splitting plugin \cite{wagner2017}. 
Within Fiji we wrote a macro based on the following processes; first the background was subtracted followed by the application of a Gaussian Blur filter. Before binarizing the image by applying a (auto-)threshold we enhanced the local contrast using CLAHE. Then we applied dilation as a morphological filter. Eventually, we used the ellipse splitting plugin to detect the droplets (excluding the ones on the side), from which we extracted the Feret's Diameter and the mean intensity. Intensities were measured from the unprocessed images. This workflow worked well for our system, as shown in Sup.~Fig.~12.
\begin{figure}
\centering
\includegraphics[width=\columnwidth]{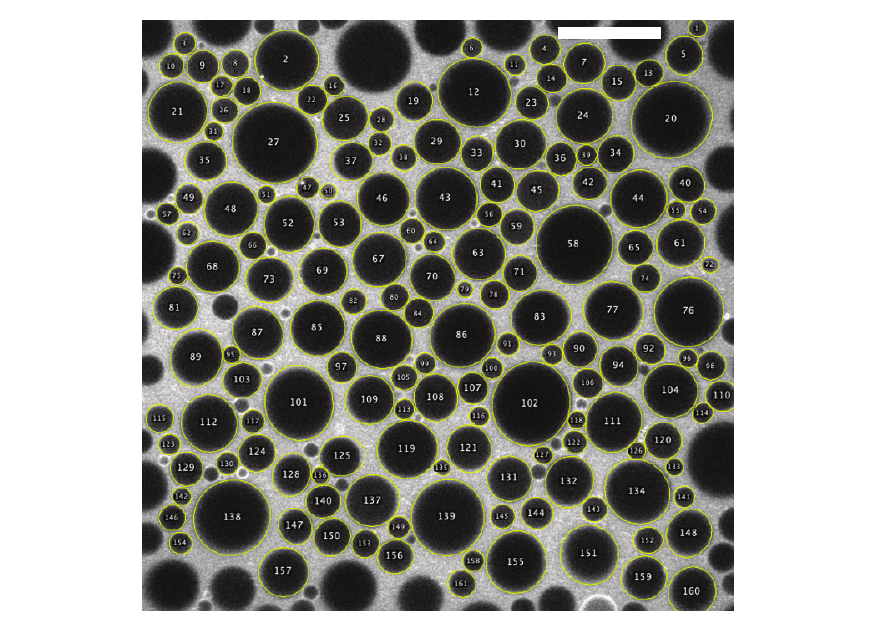}
\caption{Image analysis based on Fiji. The image processing workflow as described above detects the droplets well. The example shown here is for the emulsion stabilized with 1\% SDS, at $t_{\mathrm{0}}$. Scale bar is $20\,\upmu \mathrm{m}$.} 
\label{fig:FigS10}
\end{figure}
\newpage
\bibliography{Refs}

\end{document}